\documentclass[prb,aps,superscriptaddress,twocolumn,notitlepage,showpacs,nofootinbib]{revtex4-1}
\usepackage{latexsym}
\usepackage{amsmath, dsfont, physics}
\usepackage{amssymb,latexsym}
\usepackage{graphicx}
\usepackage{caption}
\usepackage{subfigure}
\usepackage{float}
\usepackage{mathrsfs}
\usepackage{color}
\usepackage{txfonts}
\usepackage[justification=centering,
            format=plain]{caption}

\renewcommand{\raggedright}{\leftskip=0pt \rightskip=0pt plus 0cm}

\begin{document}

\title{Anti-PT-symmetry-enhanced interconversion between microwave and optical fields}
\author{Debsuvra Mukhopadhyay}
\email{debsosu16@tamu.edu}
\affiliation{Institute for Quantum Science and Engineering, Texas A$\&$M University, College Station, TX 77843, USA}
\affiliation{Department of Physics and Astronomy, Texas A$\&$M University, College Station, TX 77843, USA}

\author{Jayakrishnan M. P. Nair}
\email{jayakrishnan00213@tamu.edu}
\affiliation{Institute for Quantum Science and Engineering, Texas A$\&$M University, College Station, TX 77843, USA}
\affiliation{Department of Physics and Astronomy, Texas A$\&$M University, College Station, TX 77843, USA}

\author{Girish S. Agarwal}
\affiliation{Institute for Quantum Science and Engineering, Texas A$\&$M University, College Station, TX 77843, USA}
\affiliation{Department of Physics and Astronomy, Texas A$\&$M University, College Station, TX 77843, USA}
\affiliation{Department of Biological and Agricultural Engineering, Texas A$\&$M University, College Station, TX 77843, USA}

\date{\today}

\begin{abstract}
The intrinsic dissipation of systems into a shared reservoir introduces coherence between two systems, enabling anti-Parity-Time (anti-PT) symmetry. In this paper, we propose an anti-PT symmetric converter, consisting of a microwave cavity coupled dissipatively to a ferromagnetic sphere, which supports significant improvements in the conversion efficiency when compared to coherently coupled setups. In particular, when only the ferrite sample is driven, the strong coherence induced by the vacuum of the mediating channel leads to much stronger enhancements in the intended conversion. The enhancement is an inalienable artifact of the emergence of a long-lived, dark mode associated with a quasi-real singularity of the hybrid system.  In addition, we observe considerable asymmetry in the efficiencies of microwave-to-optical and optical-to-microwave conversions, in spite of the symmetrical structure of the trilinear optomagnonic coupling stimulating both the transduction phenomena. The nonreciprocity stems from the intrinsic asymmetry in the couplings of the microwave and optical fields to the cavity-magnon network as well as the phase coupling entailed by the spatial separation. 
\end{abstract}

\maketitle

\section{Introduction}
Nonlinear three-wave interactions constitute the bedrock of several exotic nonlinear optical phenomena, including stimulated Raman and Brillouin scattering \cite{boyd2020nonlinear}, first observed by Woodbury \textit{et al.} \cite{woodbury1962ruby} and later explained by Garmier \textit{et al} \cite{PhysRevLett.11.160} and Bloembergen and Shen \cite{PhysRevLett.12.504,PhysRev.137.A1787} employing a coupled wave formalism. Quantum mechanically, such an interaction can be understood as the conversion of an incident pump quantum into two energy-conserving daughter quanta and vice versa. However, such exchanges do not require all fields to be optical or lie in overlapping frequency domains. For instance, in the last few decades, such three-mode interactions have been widely investigated in the context of the interconversion between optical and microwave fields, both of which are pivotal to an efficient information processing network. A coherent reversible transduction of signals between microwave and optical frequencies can leverage the strengths of optical signals, including low-loss transmission, long-time memory and low thermal occupancy, while simultaneously facilitating control over the electrical system with the application of microwave signals. A variety of experimental systems have been explored, including, for example, optomechanical systems \cite{PhysRevA.88.013815, PhysRevA.87.031802, bagci2014optical, andrews2014bidirectional} cold atoms \cite{hafezi2012atomic, verdu2009strong}, spins \cite{PhysRevLett.102.083602, marcos2010coupling}, trapped ions \cite{williamson2014magneto, fernandez2015coherent} and electro-optic \cite{tsang2010cavity, tsang2011cavity}. A conversion efficiency close to 10$\%$ between microwave signals of a few GHz into optical domain was demonstrated in an experiment \cite{andrews2014bidirectional} using optomechanical systems.

 Recently, the interconversion between microwave and optical waves was demonstrated in a hybrid cavity-magnonic setup by coherently coupling a microwave cavity to the collective spin excitations in a yttrium iron garnet (YIG) sample through the Purcell effect and by exploiting a trilinear Faraday interaction between two optical modes and a Kittel mode \cite{PhysRevB.93.174427}. Subsequently, Ihn \textit{et al.} generalized this result to the multimode case by accounting for both the Kittel mode and a higher-order space-varying magnetostatic mode \cite{PhysRevB.102.064418}. The primary challenge posed by these setups is the weakness of the nonlinear parametric interaction between the optical modes and the magnonic frequencies as they lie in widely disjoint regimes. While much progress has been accomplished with regard to improving the conversion efficiency in optomechanical configurations or by coupling an erbium-doped crystal \cite{PhysRevA.92.062313} to both a microwave and an optical cavity, mitigating the adverse impact of weak optomagnonic couplings remains a central difficulty. However, all these studies were based on dispersively coupled systems, while the recent years have seen a flourishing of interest in anti-PT symmetry achieved via the engineering of dissipative couplings. While coherent coupling stems from the spatial overlap between two modes, a dissipative type of coupling \cite{PhysRevLett.123.227201} can be engineered by the inclusion of a shared reservoir coupled independently to the two modes. Subsidiary to this development is the proliferation of experimental pursuits in cavity-magnonics \cite{PhysRevLett.111.127003,PhysRevLett.113.083603,PhysRevLett.113.156401, PhysRevLett.114.227201, zhang2015magnon, tabuchi2015coherent, chumak2015magnon, zhang2016cavity, PhysRevLett.116.223601, PhysRevLett.118.217201, zhang2017observation, PhysRevLett.120.057202, lachance2019hybrid, PhysRevB.102.104415,PhysRevB.103.224401}, courtesy of the high spin-density and low dissipation rates of YIGs. The applications of cavity-magnonics include, but are not restricted to, quantum and classical sensing \cite{PhysRevResearch.2.013031,lachance2020entanglement,PhysRevLett.125.117701,PhysRevLett.126.180401}, non-reciprocity \cite{PhysRevLett.123.127202,PhysRevApplied.13.044039}, multistability \cite{PhysRevB.102.104415} and many more. Some experiments have probed dissipative couplings as well \cite{PhysRevLett.121.137203, PhysRevB.99.134426}. 

In an earlier work, a scheme to engineer an augmented response to weak nonlinear perturbations was expounded in a dissipatively coupled cavity-YIG apparatus by exploiting the anti-PT symmetry of the configuration \cite{PhysRevLett.126.180401}. In view of the immense potential of anti-PT symmetry, we hereby extend its application to the problem of interconversion between microwave and optical fields in the specific context of cavity-magnonics. Our key observation is that the anti-PT symmetry \cite{peng2016anti,wu2014non,wu2015parity,wang2016optical,jiang2019anti,konotop2018odd,choi2018observation,li2019anti,yang2020unconventional,wen2020observation} allowed by dissipative frameworks offers a noticeable improvement in efficiency. In fact, when only the YIG sphere is driven, the same arrangement will enable improvements by a few orders of magnitude for strong dissipative couplings in otherwise low-loss environments. This is an upshot of the emergence of a long-lived eigenmode induced by the coherence. Furthermore, owing to the dissipative coupling between the two modes, we observe a strong asymmetry in the efficiencies of microwave-to-optical and optical-to-microwave conversions. Note that, in general, it is challenging to engineer non-reciprocity \cite{PhysRevX.5.021025,verhagen2017optomechanical,nassar2020nonreciprocity,wang2021induced} in physical systems; however, in waveguide-integrated photonic devices, the reservoir-mediated phase coupling naturally brings in nonreciprocal attributes. 

The structure of the paper is summarized below: Following a concise review of the theoretical formulation underpinning the conversion model in Sec. II, we explain the two possible schemes vis-\`{a}-vis the microwave-to-optical transduction mechanism in Sec. III. Finally, in Sec IV, we derive the efficiency of the reverse conversion, \textit{viz.} optical to microwave, and demonstrate the nonreciprocity of the two conversion pathways.



\section{Theoretical Conversion Model}\label{sec1}
We consider a hybrid cavity-magnonic model in which a rectangular microwave cavity and a YIG interacts dissipatively via a one-dimensional (1D) microwave transmission line. Unlike the previously explored scenario where the YIG is wedged inside the cavity resonator to enhance the coherent coupling between them through the Purcell effect \cite{PhysRevB.93.174427, PhysRevB.102.064418}, the interposing waveguide here acts as the mediator of a long-range coupling between the two. For the purpose of the conversion process, the YIG sphere is evanescently coupled to an ancillary optical fiber withal. The transduction mechanism could be initiated by the application of a microwave drive at a frequency $\omega_{\mu}$, duly complemented by an intense laser drive at a frequency $\Omega_{0}$ sent along optical fiber targeting the YIG sample. The converse mechanism of optical-to-microwave transfer proceeds via the administration of two orthogonally polarized optical inputs at frequencies $\Omega$ and $\Omega_0$ respectively, such that the frequency discord $\abs{\Omega-\Omega_0)}$ is closely resonant with a Kittel mode frequency, which lies in the microwave regime. 

At the outset, we overview a first-principle description of the exchange dynamics enabling the mode conversions. In the most general case, the full Hamiltonian $\mathcal{H}=\mathcal{H}_s+\mathcal{H}_{\text{micro}}+\mathcal{H}_{\text{optical}}$ of the driven cavity-YIG system would comprise of the following contributions:
\begin{align}
\mathcal{H}_s=&\hbar \omega_aa^{\dagger}a-\hbar\gamma_{e}B_{0}S_{z},\notag\\
\mathcal{H}_{\text{micro}}=&\mathcal{H}_{\text{micro}}^{(a)}+\mathcal{H}_{\text{micro}}^{(m)},\notag\\
\mathcal{H}_{\text{optical}}=&\int_{0}^{T}dt \: \hbar g M_{x}(t)s_{x}(t)c.
\end{align}
Here, $\mathcal{H}_s$ represents the free Hamiltonian of the cavity-YIG composite, where $a$ ($a^{\dagger}$) denotes the annihilation (creation) operator of the cavity, $B_{0}$ the applied bias magnetic field and $S_{z}$ the collective spin operator of the YIG along the z direction, and $\gamma_{e}$ is the gyromagnetic ratio. The Hamiltonian $\mathcal{H}_{\text{micro}}$ captures the typical interplay between incident microwave photons and the cavity-YIG network, with the first term $\mathcal{H}_{\text{micro}}^{(a)}$ encapsulating the coupling of the microwave drive to the cavity. The second term $\mathcal{H}_{\text{micro}}^{(m)}=-\hbar \gamma_{e}\vec{S}.\vec{B}_{\text{micro}}$ characterizes the exchange interaction between a large number of spins in the YIG and the local microwave field $\vec{B}_{\text{micro}}$. The specific expressions for these two contributions would be introduced in due course based on contextual considerations about the incident fields. In addition, two linearly polarized optical field modes selectively couple to the ferromagnetic material which is magnetized along the z direction. The parameter $T$ corresponds to the Faraday interaction time, $g$ is the coupling strength between the optical field modes and the ferromagnetic sample, $c$ is the velocity of light, $M_{x}$ and $s_x$ denote the x-components of the magnetization and the relevant stokes operator respectively. Owing to the circular birefringence of the medium, the linearly polarized light rotates with components along the perpendicular directions, producing the trilinear interaction $\mathcal{H}_{\text{optical}}$. This phenomenon is known as the Faraday effect, generating optical sidebands displaced by $\omega_m=\gamma_e B_0$ on either flank of the $\Omega_0$-band, and thereby, converting those incident microwave photons into optical photons. By the same token, the application of two orthogonally polarized optical beams stimulates the production of a microwave field oscillating at a frequency equaling the difference between the two optical frequencies, a phenomenon commonly known as the inverse Faraday effect. These two phenomena constitute the physical foundation of the conversion scheme.

\begin{figure*}
\captionsetup{justification=raggedright,singlelinecheck=false}
\begin{center}
\includegraphics[scale=0.85]{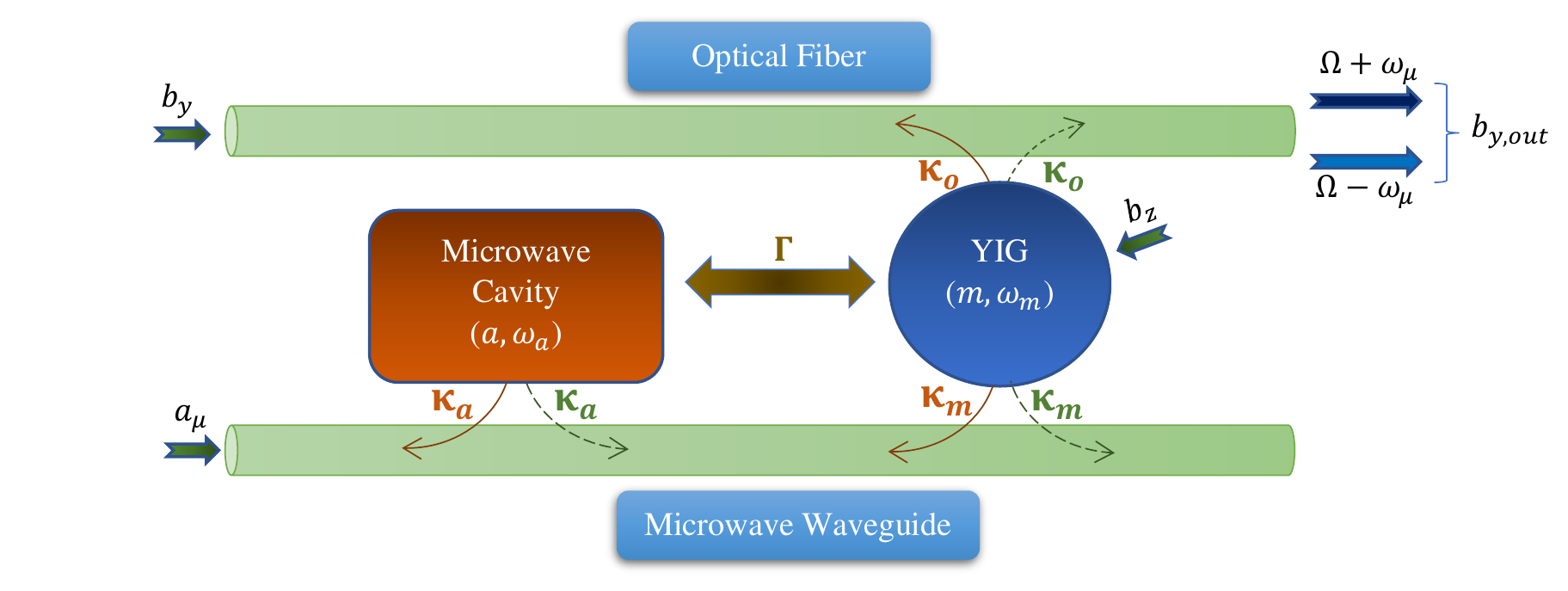}
\end{center}
\caption{First scheme for microwave-to-optical conversion. The bottom microwave waveguide couples to both the cavity and the YIG sphere, whereas the itinerant vacuum mode $b_y$ in the top fiber, polarized along $y$-axis, addresses the collective magnonic excitation. A microwave signal $a_{\mu}$ is launched through the bottom channel, in conjunction with an intense optical laser drive $b_z$, polarized along the $z$-axis and shone on the YIG. This induces both the Stokes and the anti-Stokes optical sidebands at the output port of the optical fiber.}
\label{f0}
\end{figure*}

YIG is a ferrite compound having a high ferric iron (Fe$^{3+}$) density which is approximately $4.22\times 10^{27}$m$^{-3}$, diameter $d=1$ mm and total number of spins $N\approx 10^{18}$. Employing the Holstein-Primakoff transformation in the high-spin-density limit, we can recast the raising and lowering spin operators into $S^+=\sqrt{5N}m, \ S^-=\sqrt{5N}m^{\dagger}$, where $\{m, m^\dagger\}$ follow the Bosonic algebra. This reduces the Hamiltonian of the magnons effectively to $\hbar\omega_mm^{\dagger}m$, with $\omega_m = \gamma_{e} B_{0}$. The magnetostatic Kittel mode being a spatially uniform mode, the magnetization has a simple form in terms of the collective spin operator, $M=\gamma_{e}S/V$, where $V$ is the volume of the YIG sample. Further, the Stokes operator $s_{x}$ of the input optical field can be expressed as $s_{x}=-i[b_{z}^{\dagger}b_{y}-b_{z}b_{y}^{\dagger}]$, with $b_{y}(t)$ and $b_{z}(t)$ representing two optical field operators along the y and z directions respectively. Now, since $b_{z}(t)$ is a strongly driven optical mode, we switch to a classical treatment of this mode whereby $b_{z}(t)$ is superseded by the c-number $\sqrt{\frac{P_0}{\hbar \Omega_0}}e^{-i(\Omega_0 t-\theta)}$, where $P_0$ is the input laser power and $\theta$ is the phase acquired by the wave in its transit across the fiber from the cavity to the YIG. Finally, assuming a clear distinction between the timescales of the interaction and the magnonic oscillation, i.e. $T\ll 1/{\omega_m}$, we can tailor the optomagnonic interaction in (1) into 
\begin{align}
\mathcal{H}_{\text{optical}}=-i\hbar\sqrt{\kappa_o}(m+m^{\dagger})[b_ye^{i\Omega_0t}-b_y^{\dagger} e^{-i\Omega_0t}],
\end{align}
where $\kappa_{o}=\frac{g^2 c^2 T^2 N P_0}{8 V^2 \hbar \Omega_{0}}$ is the opto-magnonic coupling strength \cite{PhysRevB.93.174427}, and a redefinition $b_y\rightarrow b_ye^{i\theta}$ has been used to eliminate the $\theta$-dependence.
The dissipative non-Hermitian coupling between the cavity and the YIG has to be formalized at the level of a master equation for the system by treating the common waveguide as a thermal bath. We can obtain the master equation for the density operator by adiabatically eliminating the slowly evolving waveguide degrees of freedom. A rigorous derivation of the same appears in the Appendix.

\section{Transduction from Microwave to Optical Field}\label{sec2}

Apropos of the microwave-to-optical  conversion process, the magnons in mode $m$ are driven by a classical microwave drive frequency $\omega_{\mu}$, while concurrently interfacing with an optical waveguide that delivers and channels the optical component $b_y$. In 1(a), the input microwave drive is launched through the bottom fiber, while in 1(b), it directly impinges on the YIG sample. The Faraday interaction $\mathcal{H}_{\text{optical}}$, prompted by the pump laser $b_z$ at the frequency $\Omega_0$ sent along the optical channel, entails the Stokes and anti-Stokes sidebands  relative to $\Omega_0$ in the optical spectrum. The indirect coupling $\Gamma$ engineered between the cavity and the magnons through the interceding transmission line consolidates the coupling of the YIG to the microwave photons via the Purcell effect. The sideband outputs can be empirically reconstructed from the amplified, beat-down, heterodyne signal between the carrier and the side modes. In principle, there could be two possible schemes of effecting the conversion: (i) by guiding the microwave drive field along the microwave fiber, or (ii) by directly shining a maser on the YIG. In what follows, we develop explicit results for the two schemes separately. \\

\subsection{Scheme 1: Microwave field launched through the waveguide}
When the microwave field is launched through the transmission line, the magnons encounter a phase-translated input relative to the intracavity field, which promptly simplifies the components in $\mathcal{H}_{\text{micro}}$ to
\begin{align}
\mathcal{H}_{\text{micro}}^{(a)}&=-i\hbar \sqrt{\kappa_a}[a^{\dagger}{a}_{\mu}e^{-i\omega_{\mu}t}-aa_{\mu}^{\dagger}e^{i\omega_{\mu}t}],\notag\\
\mathcal{H}_{\text{micro}}^{(m)}&=-i\hbar \sqrt{\kappa_m}[m^{\dagger}{a}_{\mu}e^{-i(\omega_{\mu}t-\phi)}-ma_{\mu}^{\dagger}e^{i(\omega_{\mu}t-\phi)}],
\end{align} 
where $\kappa_{a}$ and $\kappa_{m}$ symbolize the leakage rates of the cavity and the magnons respectively into the interfacing microwave line, and $a_{\mu}$ is the microwave drive. Fig. 1 portrays the overarching ladder network. It is convenient to redefine the system variables as $\tilde{a}=ae^{i\omega_{\mu}t}$, $\tilde{m}=me^{i\omega_{\mu}t}$, and, depending on whether we consider the Stokes or anti-Stokes process, $\tilde{b}_y=b_ye^{i(\Omega_0-\omega_{\mu})t}$ or $\tilde{b}_y=b_ye^{i(\Omega_0+\omega_{\mu})t}$. The operation dispels the explicit time-dependence in $\mathcal{H}$, when the fast-oscillating terms are expunged. The first process pertains to a parametric-amplification type process, whereas the second is associated with a beam-splitter type interaction. We now develop the Langevin equations for the two respective processes, to the lowest order in the optomagnonic coupling rate $\sqrt{\kappa_0}$. For notational simplicity, we drop the tildes over the mode variables.

\begin{figure}
 \captionsetup{justification=raggedright,singlelinecheck=false}
 \centering
   \includegraphics[scale=0.5]{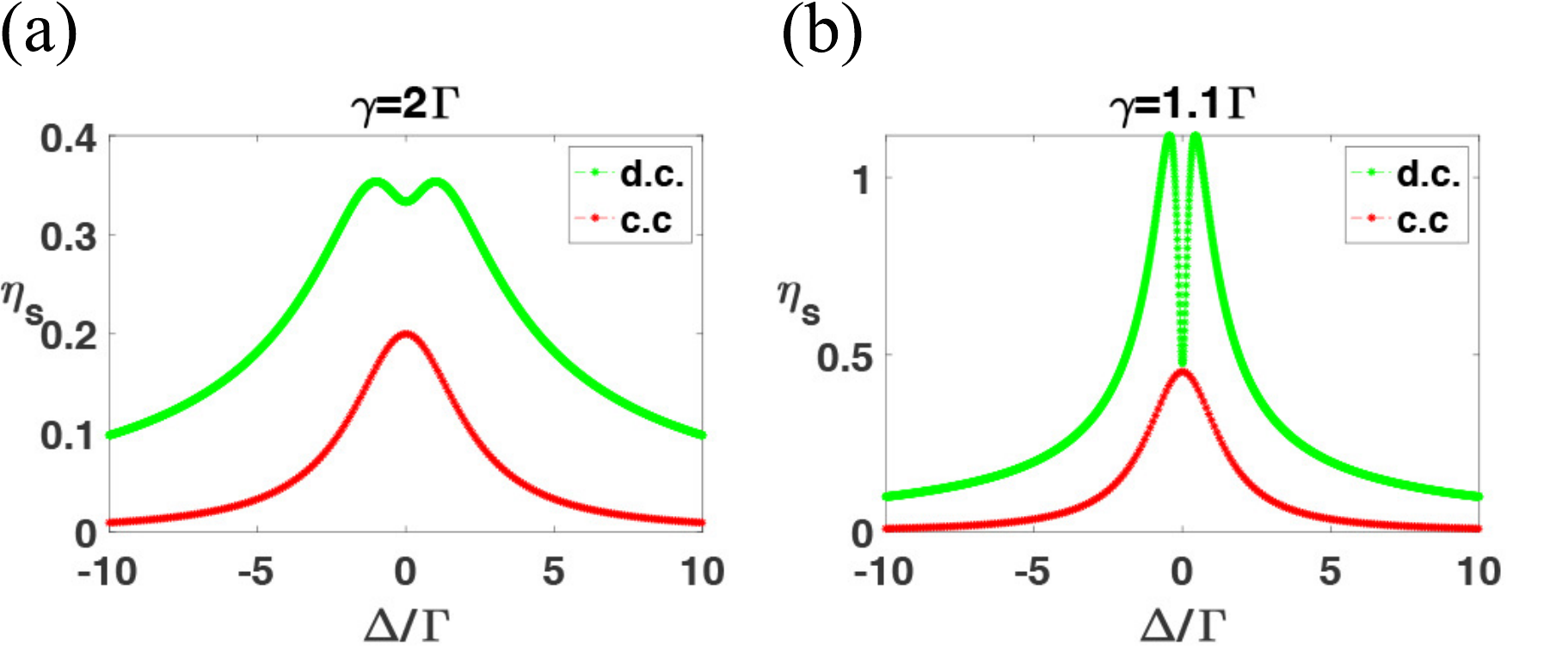}
\caption{Stokes conversion efficiencies for the microwave-to-optical conversion pertaining to scheme 1, are plotted under anti-PT symmetric conditions and comparable couplings $\Gamma=g$ in both dissipative and coherent setups. In (a), we impose $\gamma=2\Gamma$ and in (b), we consider $\gamma=1.1\Gamma$. The absolute efficiencies have been scaled up by a factor of $S=1.4\cross 10^5$ in both (a) and (b), for $\Gamma\approx\pi\cross 25$ MHz and $\Gamma_0\approx \pi\cross 0.3$ mHz.}
\label{sch}
\end{figure}

\textit{Stokes output:} Pursuant to the master equation derived in Appendix \ref{Appendix:a}, the mean-value dynamics of the cavity and the magnon modes reduces, in this case, to
\begin{align}
\dot{a}&=-(i\Delta_a+\gamma_a)a-\Gamma e^{i\phi}m-\sqrt{\kappa_a}a_{\mu},\notag\\
\dot{m}&=-(i\Delta_m+\gamma_m)m-\Gamma e^{i\phi}a-\sqrt{\kappa_m}a_{\mu}e^{i\phi}+\sqrt{\kappa_o}b_y^{\dagger}.
\end{align}
where $\gamma_a=\kappa_a+\tilde{\kappa}_a$ and $\gamma_m=\kappa_m+\tilde{\kappa}_m+\kappa_{o}\approx \kappa_m+\tilde{\kappa}_m$ encompass the dissipative effects. The decay rate $\tilde{\kappa}_a$ ($\tilde{\kappa}_m$) stands for the intrinsic damping of the cavity (magnon) mode into the non-waveguide modes. Based on the input-output formulation, the output Stokes wave would be accorded by the relation
\begin{align}
b_{y,\text{out}}=-\sqrt{\kappa_o}m^{\dagger},
\end{align}
since the input mode $b_y$, in this case, can be taken to be the thermal vacuum. Eq. (4) can be condensed into the form
\begin{align}
\dot{X}=-i\mathcal{M}X-\mathcal{F}-\sqrt{\kappa_o}\mathcal{G},
\end{align}
where $X=(a\hspace{2mm}m)^T$, $\mathcal{F}=a_{\mu}(\sqrt{\kappa_a}\hspace{2mm}\sqrt{\kappa_m}e^{i\phi})^T$, $\mathcal{G}=(0\hspace{2mm}b_y^{\dagger})^T$, and $\mathcal{M}=\begin{pmatrix}
\Delta_a-i\gamma_a & -i\Gamma e^{i\phi}\\
-i\Gamma e^{i\phi} & \Delta_m-i\gamma_m\\
\end{pmatrix}$. If $b_{y,\text{out}}$ has to be solved to the lowest (here, linear) order in $\sqrt{\kappa_0}$, it follows that the long-time solution to $X$ could be obtained by dropping the final term in (6), leading to
\begin{align}
X=i\mathcal{M}^{-1}\mathcal{F}.
\end{align}
Hence, the solutions to the intracavity field and the magnonic oscillation stand as
\begin{align}
a&=\frac{ia_{\mu}[\sqrt{\kappa_a}(\Delta_m-i\gamma_m)+i\sqrt{\kappa_m}e^{2i\phi}\Gamma]}{(\Delta_a-i\gamma_a)(\Delta_m-i\gamma_m)+\Gamma^2e^{2i\phi}},\notag\\
m&=\frac{ia_{\mu}e^{i\phi}[i\sqrt{\kappa_a}\Gamma+\sqrt{\kappa_m}(\Delta_a-i\gamma_a)]}{(\Delta_a-i\gamma_a)(\Delta_m-i\gamma_m)+\Gamma^2e^{2i\phi}}.
\end{align}
By virtue of Eq. (5), we educe the Stokes conversion efficiency to be
\begin{align}
\eta_{s}^{(d.c)}=\abs{\frac{b_{y,\text{out}}}{a_{\mu}}}=\abs{\frac{i\Gamma\sqrt{\kappa_o\kappa_a}+(\Delta_a-i\gamma_a)\sqrt{\kappa_o\kappa_m}}{(\Delta_a-i\gamma_a)(\Delta_m-i\gamma_m)+\Gamma^2e^{2i\phi}}},
\end{align}
For comparison, we recall the efficiency factor in a coherent environment, where the cavity and the magnons address each other by means of a Hermitian coupling $J(a^{\dagger}m+am^{\dagger})$: 
\begin{align}
\eta_{s}^{(c.c)}=\abs{\frac{b_{y,\text{out}}}{a_{\mu}}}=\abs{\frac{J\sqrt{\kappa_o\kappa_a}}{(\Delta_a-i\gamma_a)(\Delta_m-i\gamma_m)-J^2}}.
\end{align}
Since the solution to $X$ determines the Stokes sideband of the optical spectra, its efficiency of conversion from the microwave input pivots on the symmetry properties of $\mathcal{M}$. In this context, it is useful to note that anti-PT symmetry could be engineered in this bimodal framework by devising the constraints $\Delta_a=-\Delta_m=\Delta$, $\gamma_1=\gamma_2=\gamma$ and $\phi=n\pi$, with $n$ being a natural number. While, from the above expressions, it is clear that the order of magnitude does not change dramatically by switching to a dissipative configuration, it does turn out that dissipative couplings can predominantly outstrip the efficiency of coherent setups under anti-PT conditions. From Eqs. (9) and (10), the figure of merit quantified as an advantage gained through anti-PT symmetry in dissipative environments can be simplified as
\begin{align}
\dfrac{\eta_{s}^{(d.c)}}{\eta_{s}^{(c.c)}}=\bigg[\frac{\{\Delta^2+(\gamma-\Gamma)^2\}\{\Delta^2+\gamma^2+\Gamma^2\}}{\Gamma^2\{\Delta^2+\gamma^2-\Gamma^2\}}\bigg]^{1/2},
\end{align}
where $\kappa_a\approx\kappa_m$ and $\Gamma\approx J$ has been assumed. Figures 2(a) and 2(b) plot the conversion factors as well as the relative efficiencies w.r.t coherent couplings for comparable system parameters and similar coupling strengths, under the constraint of anti-PT symmetry. It is interesting to note the broadband nature of the profiles for anti-PT symmetry. For large dampings or frequency detunings in the range $\Delta\geq\Gamma$, the expression above exceeds 1, demonstrating the superiority of these couplings. In the complementary regime, the situation is not so cut and dry, and numerical means become the imperative.\\

\textit{Anti-Stokes output:} An identical calculation for the anti-Stokes process yields the same efficiency factor as above, i.e.,
\begin{align}
\eta_{as}^{(d.c)}=\abs{\frac{b_{y,\text{out}}}{a_{\mu}}}=\abs{\frac{i\Gamma\sqrt{\kappa_o\kappa_a}+(\Delta_a-i\gamma_a)\sqrt{\kappa_o\kappa_m}}{(\Delta_a-i\gamma_a)(\Delta_m-i\gamma_m)+\Gamma^2e^{2i\phi}}},
\end{align}
where the relevant input-output relation, $b_{y,\text{out}}=-\sqrt{\kappa_o}m$, has been employed. Thus, the input microwave photons of frequency $\omega_{\mu}$, upon interacting with the optical wave of frequency $\Omega_0$, are converted into output traveling light modes with frequencies $\Omega_0-\omega_{\mu}$ and $\Omega_0+\omega_{\mu}$ pertaining to the Stokes and anti-Stokes bands respectively.\\

\begin{figure*}
\captionsetup{justification=raggedright,singlelinecheck=false}
\begin{center}
\includegraphics[scale=0.85]{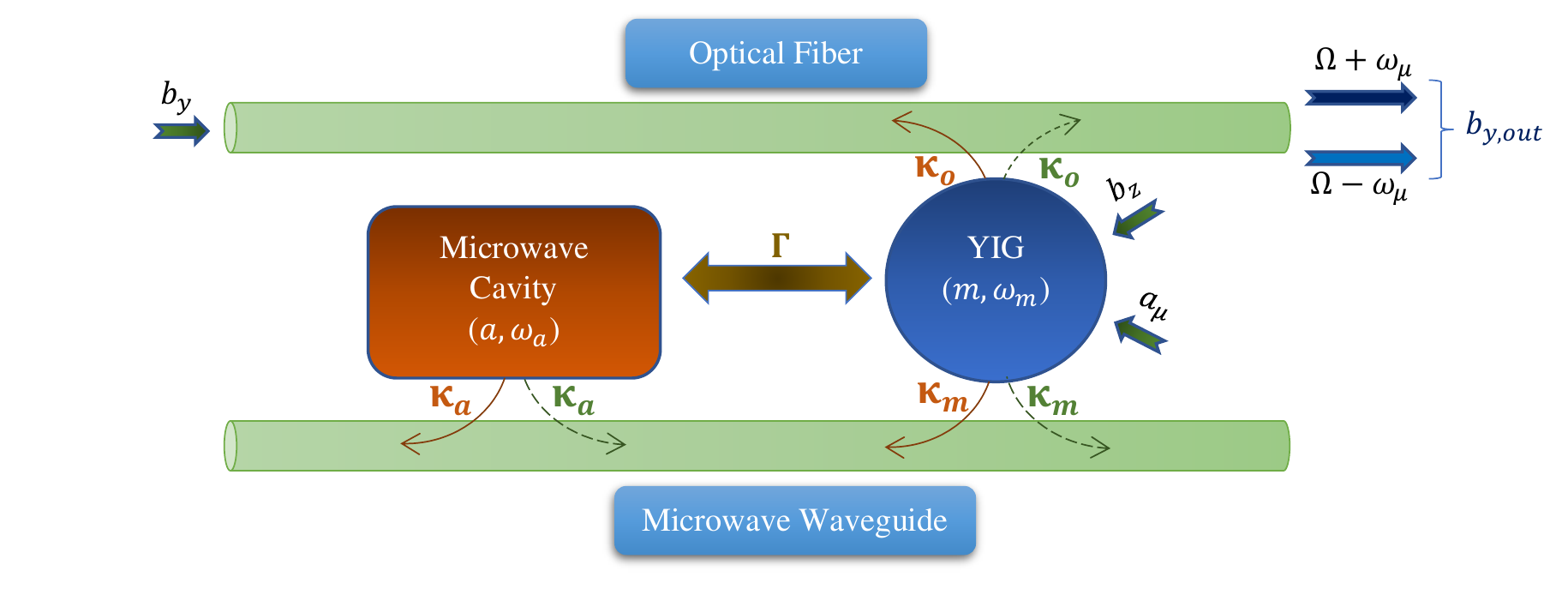}
\end{center}
\caption{Second scheme for microwave-to-optical conversion. The microwave input as well as the laser drive are targeted on the YIG sphere, generating the optical sidebands.}
\label{f0}
\end{figure*}

\subsection{Scheme 2: Direct excitation of the magnons by a microwave pump}

\begin{figure}
 \captionsetup{justification=raggedright,singlelinecheck=false}
 \centering
   \includegraphics[scale=0.4]{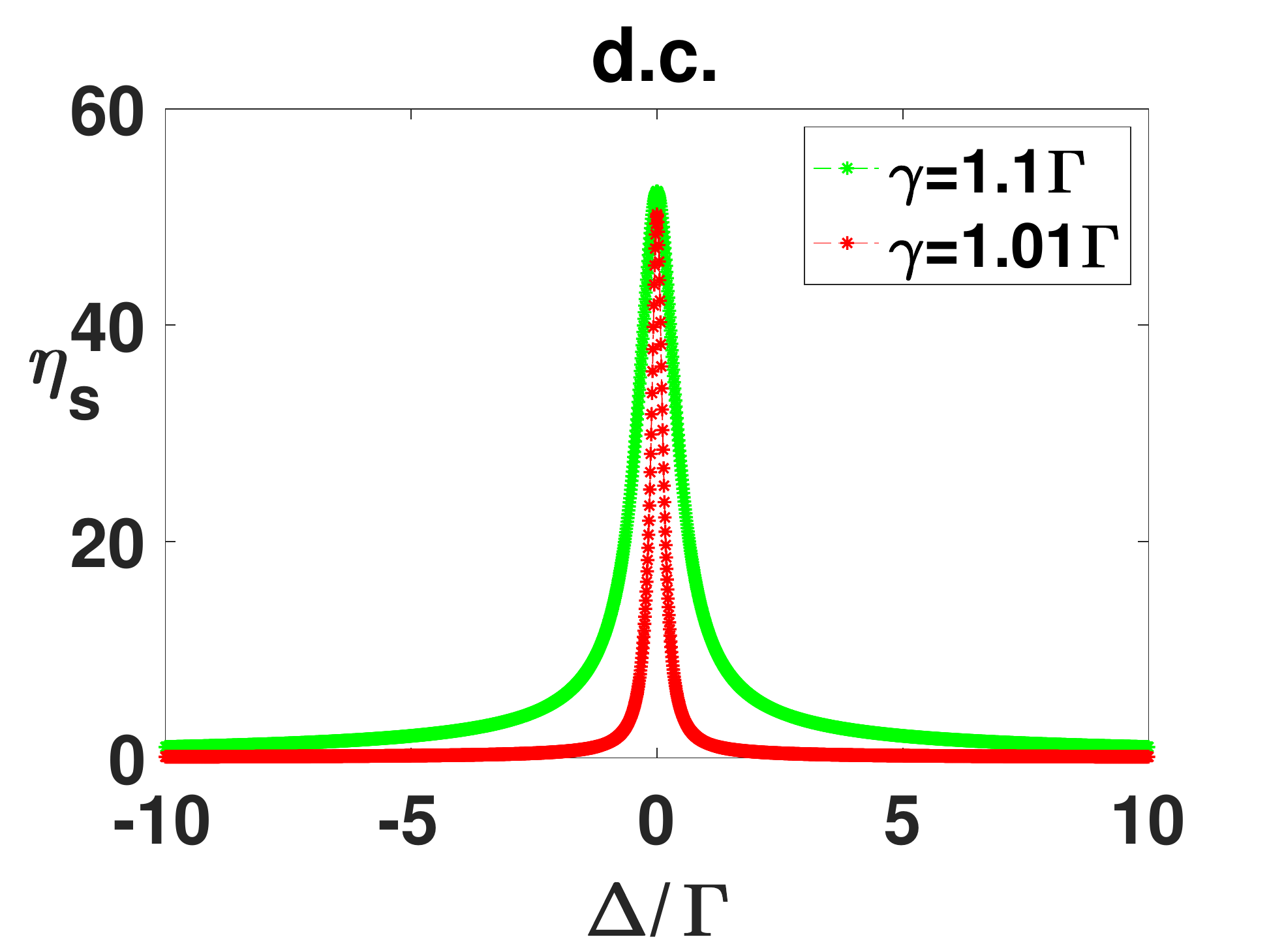}
\caption{Stokes conversion efficiencies for the scheme 2 with the chosen parameters $\gamma=1.1\Gamma$ (i.e., $\varepsilon=0.1$) and $\gamma=1.01\kappa$ (i.e., $\varepsilon=0.01$), under anti-PT symmetry. The absolute efficiencies have been scaled up by $S(\varepsilon=0.1)=1.4\cross 10^5$ and $S(\varepsilon=0.01)=1.4\cross 10^4$. Since the peaks in the two graphs are in the vicinity of each other, the proportionality in Eq. (15) is vindicated.}
\label{sch}
\end{figure}

This protocol of selectively driving the YIG sphere, as demonstrated in Fig. 3, can exploit a spectral singularity of dissipatively interacting systems in drawing out an enhanced steady-state response in the hybrid system. The vacuum of the quantized field in a reservoir induces coherence between any two systems coupled to it. When optimally strong, the coherence can push one of the poles in the linear response to the real axis, under anti-PT symmetric conditions. With the corresponding linewidth suffering a stark suppression, the resonant response shoots up, only to be regularized by intrinsic anharmonicites present in any of the modes. This feature in a two-mode dissipatively coupled system has been tailored into a convenient mechanism for sensing weak nonlinear perturbations \cite{PhysRevLett.126.180401}. Although, for our current analysis, we do not meander into the nonlinear domain, we throw light on the feasibility of ramping up the efficiency of our conversion model by operating in the neighborhood of this singularity. With this constraint, it is easy to see that the poles in Eq. (9) approach zero as $\Gamma\rightarrow\gamma$. This is possible under the circumstance when extraneous decoherence is strongly overshadowed by the coherence produced by the shared reservoir. However, around this point, the numerator also becomes small, precluding sizable enhancements in the output fields procreated via scheme 1. To mitigate the pernicious role of the numerator, we can merely resort to shining a maser beam on the YIG sample. While the microwave-YIG interaction becomes  $\mathcal{H}_{\text{micro}}^{(m)}=-i\hbar \sqrt{2\kappa_m}[m^{\dagger}{a}_{\mu}e^{-i(\omega_{\mu}t-\phi)}-ma_{\mu}^{\dagger}e^{i(\omega_{\mu}t-\phi)}]$, the cavity gets decoupled from any external driving fields. Therefore, the term $\mathcal{H}_{\text{micro}}^{(a)}$ drops out. The Langevin equations for the mode amplitudes corresponding to the Stokes process would, then, be remodeled as
\begin{align}
\dot{a}&=-(i\Delta_a+\gamma_a)a-\Gamma e^{i\phi}m,\notag\\
\dot{m}&=-(i\Delta_m+\gamma_m)m-\Gamma e^{i\phi} a-\sqrt{2\kappa_m}a_{\mu}e^{i\phi}+\sqrt{\kappa_o}b_y^{\dagger}.
\end{align}
The input-output relation pertaining to the optical mode $b_y$ remains intact, which yields the Stokes efficiency factor to be
\begin{align}
\eta_{s}^{(d.c)}=\abs{\frac{b_{y,\text{out}}}{a_{\mu}}}=\frac{\sqrt{2\kappa_o\kappa_m(\Delta_a^2+\gamma_a^2)}}{(\Delta_a-i\gamma_a)(\Delta_m-i\gamma_m)+\Gamma^2e^{2i\phi}}
\end{align}
To bring out the significance of the anti-PT symmetry, we look at the behavior of the above expression under this constraint, i.e., $\Delta_a=-\Delta_m=\Delta$, $\gamma_1=\gamma_2=\gamma$ and $\phi=n\pi$ . The efficiency is plotted in Fig. 4, which unveils a spike around the origin $\Delta/\kappa=0$. When $\gamma$ is only slightly larger than $\Gamma$, such that $\gamma=\Gamma(1+\varepsilon)$ for $\varepsilon\ll 1$, the above expression reduces, in the limit $\Delta/\Gamma\rightarrow 0$, to
\begin{align}
\eta_{s}^{(d.c)}\approx(\varepsilon/2)^{-1}(\kappa_o/\kappa_a)^{1/2}.
\end{align}
Consequently, the smallness of $\varepsilon$ has a direct bearing on the scaling up of the figure of merit characterizing the conversion scheme. For instance, when $\varepsilon$ becomes one-tenth, there is about a tenfold amplification observed in the optical output. This is evident from the comparative plot in Fig. 4, since the deviations of $\gamma$ from $\Gamma$ are respectively one and two orders of magnitude smaller than $\Gamma$. For $\varepsilon\approx 0.1$, the value of $\eta_{s}^{(d.c)}$ approximates to $0.025\%$, which signifies a remarkable improvement over the coherent setting. 

This property in a dissipatively coupled system can also be justified from the perspective of bright and dark states pertaining to radiating and non-radiating modes of the system. On defining two linearly independent modes $c_{\pm}=1/{\sqrt{2}}(a\pm m)$, we find that the mode $c_{-}$ evolves as $\dot{c}_{-}=-\varepsilon\Gamma c_{-}-i\Delta c_{+}-\sqrt{\kappa_m}a_{\mu}+\sqrt{\kappa_0}b_y^{\dagger}$. Thus, the mode $c_-$ acts as a long-lived mode, akin to a dark state, as $\varepsilon$ becomes small. The other mode $c_+$ decays significantly faster at the rate of $2\Gamma$. The emergence of a dark state explains the coherent buildup in the output signal around the singularity $\varepsilon=0$.

\section{Optical-to-microwave conversion: Nonreciprocity due to dissipative coupling}\label{sec3}
\begin{figure}
 \captionsetup{justification=raggedright,singlelinecheck=false}
 \centering
   \includegraphics[scale=0.48]{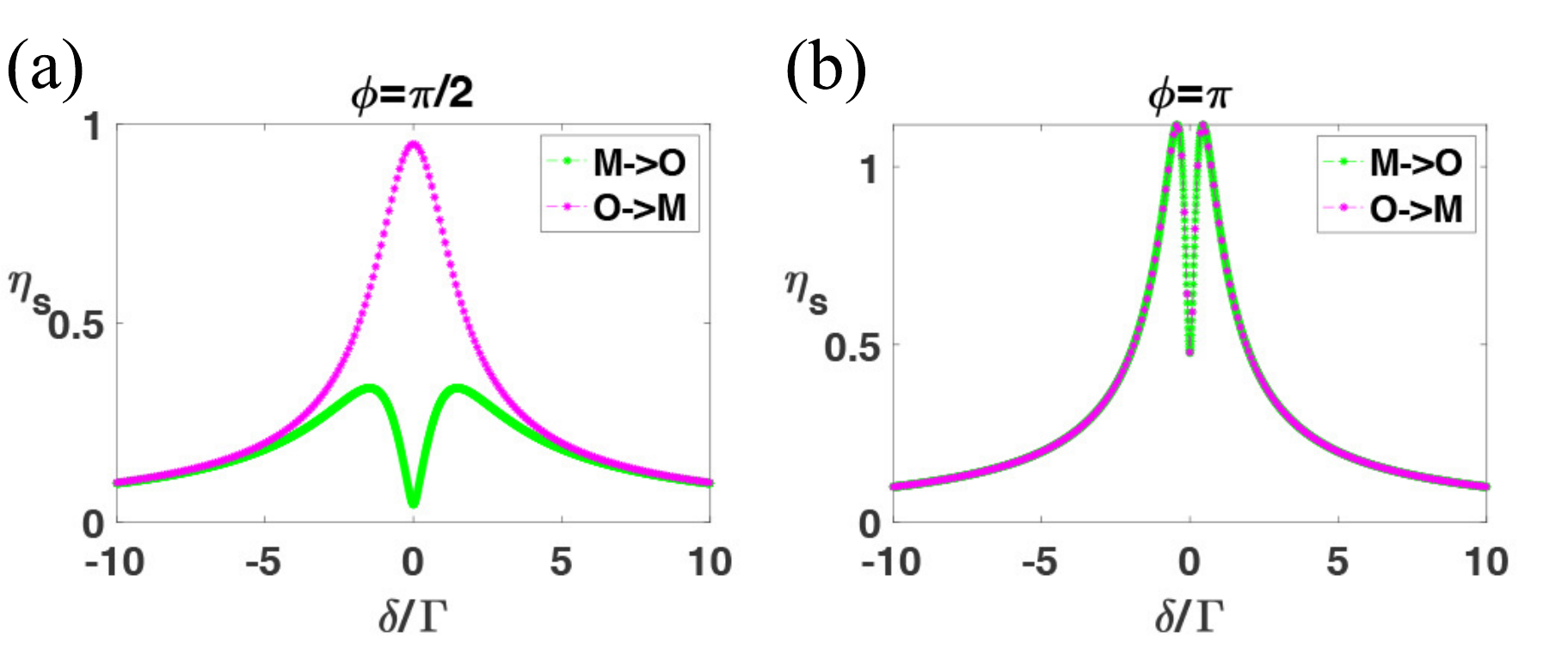}
\caption{Nonreciprocal Stokes' conversion efficiencies (M: Microwave, O: Optical) pertaining to the scheme 1, for the system parameters $\Delta_a=-\Delta_m=\delta$ (for M$\rightarrow$O) and $\delta_a^{(+)}=-\delta_m^{(+)}=\delta$ (for O$\rightarrow$M), plotted against $\delta$for two sets of the phase separation $\phi$. The conversion is, however, reciprocal when $\phi$ is an integer multiple of $\pi$. The M$\rightarrow$O graphs refer to Eq. (9), and the O$\rightarrow$M graphs refer to Eq. (19).}
\label{sch}
\end{figure}
The reverse procedure of light getting transformed into microwave photons is made feasible by the inverse Faraday effect. The initial conditions are now translated into injecting two copropagating phase-coherent laser inputs ($b_y$ and $b_z$) along the optical transmission line, with the relevant frequencies in the optical domain and separated by a microwave frequency. If the difference frequency coincides with a Kittel mode frequency, the resonance imparts an oscillatory magnetization to the magnons in the YIG, which, in turn, elicit traveling microwave photons at this difference frequency. With the plane-wave ansatz $b_y=\sqrt{\frac{P}{\hbar\Omega}}e^{-i\Omega t}$, where $P$ is the drive power, we can derive the induced microwave output at frequency $\omega_{+}=\Omega_0-\Omega$ (Stokes scattering) or $\omega_-=\Omega-\Omega_0$ (anti-Stokes scattering), depending on whichever is positive, by solving the dynamical equations in time domain. On rotating the variables as $m\rightarrow me^{i\omega_{\mu}t}$ and $a\rightarrow ae^{i\omega_{\mu}t}$, and defining the input amplitude $\beta=\sqrt{\frac{P}{\hbar\Omega}}$, we reduce the evolution equations as
\begin{align}
\dot{a}&=-(i\delta_a^{(\pm)}+\gamma_1)a-\Gamma e^{i\phi} m,\notag\\
\dot{m}&=-(i\delta_m^{(\pm)}+\gamma_2)m-\Gamma e^{i\phi}a\pm\sqrt{\kappa_o}\beta,
\end{align}
where the detunings $\delta_a^{(\pm)}=\omega_a-\omega_{\pm}$ and $\delta_m^{(\pm)}=\omega_m-\omega_{\pm}$ have been introduced, and the rapid oscillations neglected. Since no classical microwave drive needs to be applied, the corresponding interaction term has been dropped. It is now a simple exercise to evaluate the scattered microwave field in the long-time limit. Clearly, the long-time solutions $X_{\pm}$ would correspond to fresh microwave oscillations at $\omega_{+}$ or $\omega_{-}$, as the case may be, as an artifact of the nonlinear interaction between the two optical beams and the driven magnons. Eq. (16) yields the solutions
\begin{align}
X_{\pm}=\mp i\sqrt{\kappa_0}\beta\mathcal{H}^{-1}\begin{pmatrix}
0\\
1\\
\end{pmatrix}.
\end{align}

Invoking the input-output relation for the field transmitted across the microwave waveguide, we find that
\begin{align}
a_{\text{out}}^{\pm}&=-(\sqrt{\kappa_a}e^{i\phi}a_{\pm}+\sqrt{\kappa_m} m_{\pm})\notag\\
&=\mp i\beta\bigg[\frac{i\Gamma e^{2i\phi}\sqrt{\kappa_0\kappa_a}+\sqrt{\kappa_o\kappa_m}(\delta_a^{\pm}-i\gamma_1)}{(\delta_a^{\pm}-i\gamma_1)(\delta_m^{\pm}-i\gamma_2)+\Gamma^2e^{i\phi}}\bigg].
\end{align}
This allows us to infer the conversion efficiencies for the two possible microwave bands,
\begin{align}
\abs{\frac{a_{\text{out}}^{\pm}}{\beta}}=\abs{\frac{i\Gamma e^{2i\phi}\sqrt{\kappa_o\kappa_a}+(\delta_a^{\pm}-i\gamma_1)\sqrt{\kappa_o\kappa_m}}{(\delta_a^{\pm}-i\gamma_1)(\delta_m^{\pm}-i\gamma_2)+\Gamma^2e^{2i\phi}}}.
\end{align}
where the upper sign refers to the Stokes band and the lower to the anti-Stokes generation. 
\begin{figure}
 \captionsetup{justification=raggedright,singlelinecheck=false}
 \centering
   \includegraphics[scale=0.48]{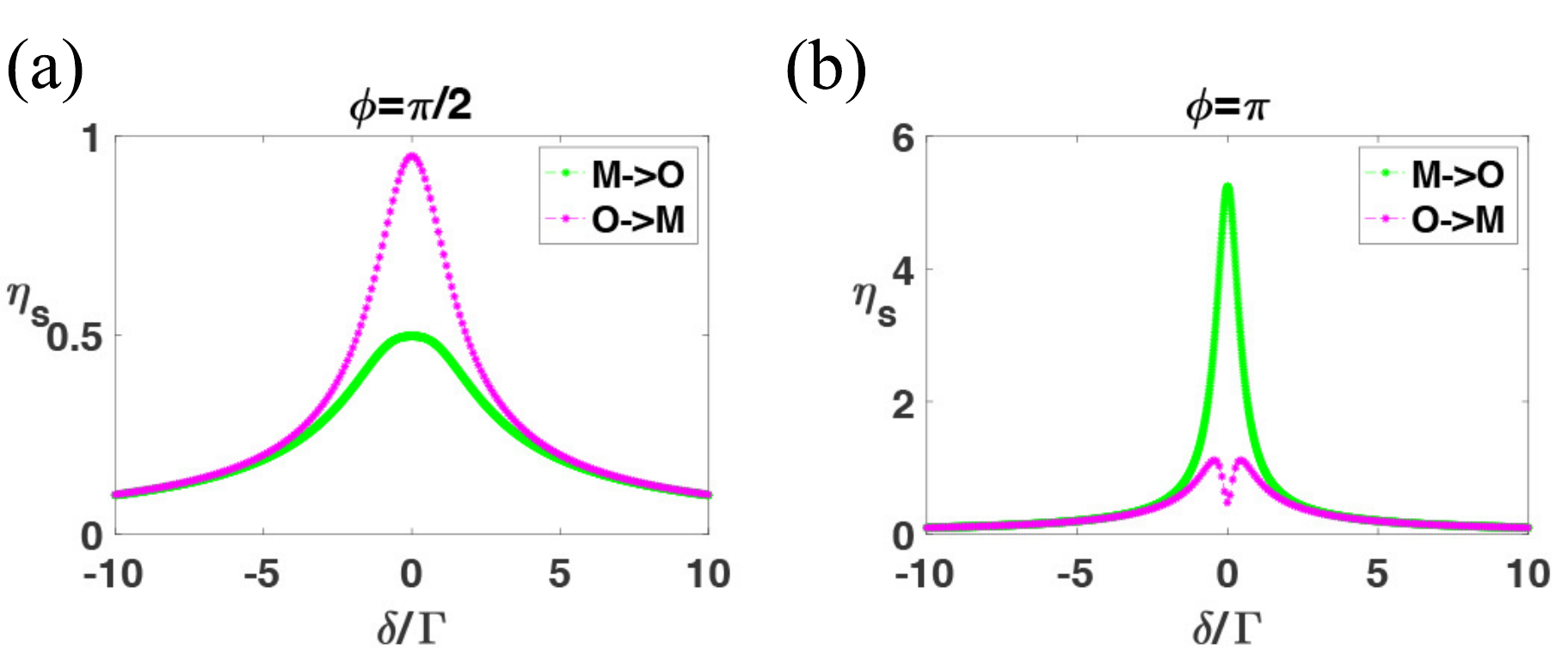}
\caption{Nonreciprocity in the Stokes' conversion efficiencies (M: Microwave, O: Optical) pertaining to scheme 2, for the system parameters $\Delta_a=-\Delta_m=\delta$ (for M$\rightarrow$O) and $\delta_a^{(+)}=-\delta_m^{(+)}=\delta$ (for O$\rightarrow$M), and $\gamma_a=\gamma_m=1.1\kappa$.The M$\rightarrow$O graphs refer to Eq. (14), and the O$\rightarrow$M graphs to Eq. (19).}
\label{sch}
\end{figure}

Strikingly, there is a structural disparity between the expression in Eqs. (9) or (14), and that in Eq. (19), indicating nonreciprocal transductions. In the case of scheme 1, for $\phi=n\pi$, the microwave-to-optical conversion efficiency and vice versa resemble each other up to a reinterpretation of the detunings. However, there is a palpable quantitative asymmetry in the two mechanisms pertaining to an arbitrary phase for an otherwise commensurate set of system parameters. The discrepancy between the two efficiency factors is even more evident in scheme 2, whereby the two conversion processes ensue with starkly disparate efficiencies, regardless of the choice of phase. Fundamental to this nonreciprocity in either scheme is the factor that microwave fields are directly or indirectly linked up with both the cavity and the YIG, whereas the optical fields interact purely with the YIG. In addition, the phase-sensitive asymmetry pertinent to scheme 1 can be traced down to the existence of the reservoir-mediated phase-coupling between the cavity and the YIG, that is a characteristic of dissipatively coupled systems. This is veritably distinct from the symmetric nature of conversion observed in coherently coupled setups, where both the conversion mechanisms unfold with equal efficiencies. Note that this nonreciprocity in the conversion mechanisms exists in spite of the symmetrical interaction between the two optical modes and the magnons, as embodied in $\mathcal{H}_{\text{optical}}$.

\section{Conclusion}\label{sec5}
In conclusion, we have demonstrated the efficient interconversion between optical and microwave fields in the context of cavity-magnonicsa dissipatively coupled optomagnonic setup. As reported in the manuscript, dissipatively coupled systems with anti-PT symmetry perform significantly better than coherent settings for comparable system parameters. We have explicated two disparate schemes for the microwave-to-optical conversion, one of which involves injecting the microwave field through the shared waveguide and the other where the ferromagnetic sample is directly subject to an external pump. While the two schemes showcase improved conversion efficiencies compared to the coherently coupled systems, the second scheme, which directly drives the magnetic sample, manifests remarkable improvements when the dissipative coupling dominates the extraneous dissipations. This superior conversion efficiency stems from the emergence of a long-lived dark mode, and consequently, the linear response suffers a tremendous boost. The transduction protocols achieved via strong phase-sensitive dissipative couplings also demonstrate strong nonreciprocity, with apparent discrepancies in the efficiencies of microwave to optical conversion and vice versa.

\section{Acknowledgements}
The authors acknowledge the support of The Air Force Office of Scientific Research [AFOSR award no FA9550-20-1-0366], The Robert A. Welch Foundation [grant no A-1943] and the Herman F. Heep and Minnie Belle Heep Texas A\&M University endowed fund.
\appendix
\section{Derivation of the master equation for a chain of emitters coupled to a waveguide}
\label{Appendix:a}
\begin{figure*}
\captionsetup{justification=raggedright,singlelinecheck=false}
\begin{center}
\includegraphics[scale=0.85]{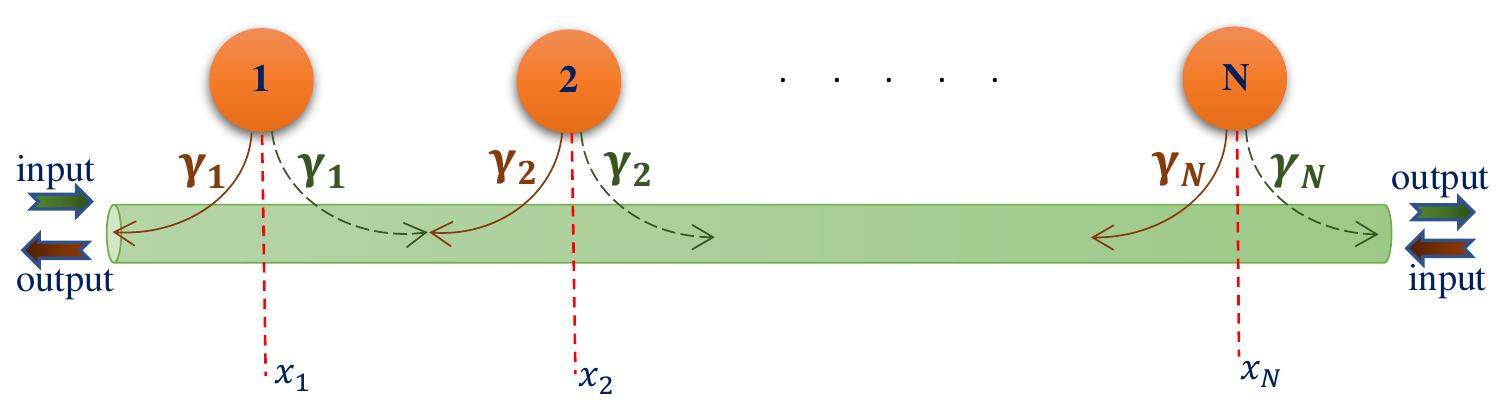}
\end{center}
\caption{An array of $N$ single-mode quantum emitters coupled to the evanescent field of a one-dimensional waveguide, with $\gamma_\alpha=\kappa_{\alpha\alpha}$ denoting the individual coupling rates. The waveguide can be adiabatically eliminated to yield the master equation of the emitter chain.}
\label{f_app}
\end{figure*}

As depicted in Fig. 7, we consider an $N$-mode system (S) interacting dissipatively through a shared one-dimensional bath (B) aligned along the $x$-axis. Keeping the model very basic, we split the net Hamiltonian $\mathcal{H}$ into three contributions given by
\begin{align}
\mathcal{H}_{S} &=\hbar\sum_{\lambda=1}^N\omega_{\lambda} c_{\lambda}^{\dagger}c_{\lambda}, \notag\\
\mathcal{H}_{B} &=\hbar\sum_{k}\omega_ka_k^{\dagger}a_k, \notag\\
\mathcal{H}_{SB} &=i\hbar\sum_{k}\sum_{\lambda=1}^2 g_{k\lambda}(a_ke^{ikx_{\lambda}}-a_k^{\dagger}e^{-ikx_{\lambda}})(c_{\lambda}+c_{\lambda}^{\dagger}),
\end{align}
where we assume that the mode $c_\lambda$ is coupled to the waveguide at the location $x=x_{\lambda}$. The coupling coefficients $g_{k\lambda}$'s are assumed to be real. Here, $\mathcal{H}_{SB}$ exemplifies a typical two-body interaction among spatially separated modes with electromagnetic field quantized in a one-dimensional geometry. We can adiabatically eliminate the reservoir degrees of freedom to obtain the master equation under the Markov approximation as
\begin{align}
\dot{\rho}_S(t)=&-\frac{i}{\hbar}[\mathcal{H}_S,\rho_S(t)]\notag\\
&-\frac{1}{\hbar^2}\int_{0}^{\infty}\dd{\tau}\Tr_B[\mathcal{H}_{SB},[\mathcal{H}_{SB}(-\tau),\rho_S(t)\rho_B]]
\end{align}
where $\mathcal{O}(-\tau)=\exp[-\frac{i}{\hbar}(\mathcal{H}_S+\mathcal{H}_B)\tau]\mathcal{O}\exp[\frac{i}{\hbar}(\mathcal{H}_S+\mathcal{H}_B)\tau]$, and $\rho_B$ is the initial state of the bath \cite{agarwal2012quantum}. Idealizing the bath to be a thermalized vacuum at zero temperature, the reservoir signatures would encoded as $\expval{a_ka_{k'}}=\expval{a_k^{\dagger}a_{k'}}=0$ and $\expval{a_ka_{k'}^{\dagger}}=\delta_{k,k'}$. Substituting these expressions into (21), we obtain $\tau-$integrals which, for the moment, can be expressed in terms of
\begin{align}
\mathcal{T}_{k\mu}^{(\pm)}=\int_{0}^{\infty}\dd\tau e^{i(\omega_{\mu}\pm\omega_k)\tau}=\pi\delta(\omega_{\mu}\pm\omega_k)+i\text{P}\frac{1}{(\omega_{\mu}\pm\omega_k)},
\end{align}
where $\text{P}(1/x)$ denotes the Cauchy Principal Value of its argument. Since $\omega_k$'s are all positive, terms of the form $\delta(\omega_{\mu}+\omega_k)$ can be stamped out. In light of these simplifications, we can compactify (21) into the form
\begin{align}
\dot{\rho}_S=-\frac{i}{\hbar}[\mathcal{H}_S,\rho_S]-\frac{1}{\hbar^2}\sum_{\alpha,\beta=1}^{N}(\mathcal{W}_{\alpha\beta}+\mathcal{W}_{\alpha\beta}^{\dagger}),
\end{align}
where a typical contribution would appear as
\begin{align}
\mathcal{W}_{\alpha\beta}=\sum_{k}g_{k\alpha}g_{k\beta}e^{ik(x_{\alpha}-x_{\beta})}\bigg[\mathcal{T}_{k\beta}^{(-)}(c_{\alpha}^{\dagger}c_{\beta}\rho_S - c_{\beta}\rho_Sc_{\alpha}^{\dagger})+\notag \\ \mathcal{T}_{k\beta}^{(+)}(c_{\alpha}c_{\beta}^{\dagger}\rho_S - c_{\beta}^{\dagger}\rho_Sc_{\alpha})\bigg].
\end{align}
We have dropped the fast-oscillating terms that go as $c_{\alpha}c_{\beta}$ or $c_{\alpha}^{\dagger}c_{\beta}^{\dagger}$ under the purview of the rotating wave approximation. Using a linearized approximation to the reservoir frequencies by letting $\omega_k\approx v_g\abs{k}$, we take the continuum limit $\sum_k\rightarrow \frac{L}{2\pi}\int\dd k$ in computing $\mathcal{W}_{\alpha\beta}$ and $\mathcal{W}_{\alpha\beta}^{\dagger}$. With the identification $\Lambda_{\pm}=\frac{1}{\pi}\text{P}\int_{-\infty}^{\infty}\dd k\frac{e^{ik(x_{\alpha}-x_{\beta})}}{\omega_{\beta}\pm v_g\abs{k}}$ and the assumption that the coupling to the modes is independent of the field's propagation direction, the cardinal intermediate relations could be codified as
\begin{align}
\int_{-\infty}^{\infty}\dd ke^{\pm ikx_{\alpha\beta}}\delta(\omega_{\beta}-v_g\abs{k})&=\frac{2}{v_g}\cos(k_{\beta}x_{\alpha\beta}),\notag\\
\Lambda_+ + \Lambda_-&=\frac{2}{v_g}\sin(k_{\beta}x_{\alpha\beta}),
\end{align}
where $x_{\alpha\beta}=\abs{x_{\alpha}-x_{\beta}}$ and $k_{\beta}=\omega_{\beta}/v_g$. Terms like $\text{P}\int_{-\infty}^{\infty}\dd k\frac{1}{\omega_{\beta}\pm v_g\abs{k}}$ get reflected as small frequency shifts in $\omega_1$ and $\omega_2$, which can be neglected. Then, collecting the like terms together in (23) and exploiting the preceding relations in (24) and (25), we obtain the full master equation for the dissipative dynamics of S:
\begin{align}
\dot{\rho}_S=-\frac{i}{\hbar}[\mathcal{H}_S,\rho_S]-\sum_{\alpha,\beta=1}^N\kappa_{\alpha\beta}(c_{\alpha}^{\dagger}c_{\beta}\rho_S-2c_{\beta}\rho_S c_{\alpha}^{\dagger}-\rho_S c_{\alpha}^{\dagger}c_{\beta})\notag \\-i\sum_{\alpha\neq \beta}\Omega_{\alpha\beta}[c_{\alpha}^{\dagger}c_{\beta}, \rho_S],
\end{align}
where the off-diagonal coefficients $\kappa_{\alpha\beta}$, for $\alpha\neq \beta$, signify dissipative couplings, while $\Omega_{\alpha\beta}$ simulate dispersive interactions. The relevant coefficients are given by
\begin{align}
\kappa_{\alpha\beta}&=\frac{g_{\alpha}^2L}{v_g}\delta_{\alpha\beta}+\Gamma_{\alpha\beta}\cos\phi_{\alpha\beta}(1-\delta_{\alpha\beta}),\notag\\
\Gamma_{\alpha\beta}&=(\kappa_{\alpha\alpha}\kappa_{\beta\beta})^{1/2}=\frac{g_{\alpha}g_{\beta}L}{v_g},\notag\\
\Omega_{\alpha\beta}&=\Gamma_{\alpha\beta}\sin\phi_{\alpha\beta},
\end{align}
where, in view of the proximity between the transition frequencies, it is assumed that $k_{\lambda}\approx k_0=\omega_0/v_g$ $\forall \lambda\in\{1,N\}$. $g_{k\alpha}\approx g_{\alpha}$, and $\{k_0,\omega_0\}$ is the central waveguide mode in the vicinity of which the linear dispersion holds valid. The phases $\phi_{\alpha\beta}$ are defined as $\phi_{\alpha\beta}=k_0x_{\alpha\beta}$, and $g_{k\lambda}$'s are taken to be $k$-independent. When $\phi_{\alpha\beta}$'s are integral multiples of $\pi$, the couplings are purely dissipative. Note that the decay parameter $\kappa_{\alpha\alpha}=\frac{g_{\alpha}^2L}{v_g}$ deduced here accounts only for the waveguide's contribution to the dynamics. When other decohering channels are considered in parallel, additional dissipative effects are tacked onto these terms.\\

\bibliography{references}

\end{document}